\documentstyle[aps,12pt]{revtex}

\begin{document}
\title{Periodic Instanton and Phase Transition in Quantum Tunneling of Spin Systems}
\author{Yunbo Zhang$^{a,b}$, Yihang Nie$^b$, Supeng Kou$^{a,b}$, Jiuqing Liang$%
^{b,a} $, H. J. W. M\"{u}ller-Kirsten$^c$, Fu-Cho Pu$^{d,a}$}
\address{$^a${\small Institute of Physics and Center for Condensed Matter Physics,\\
Chinese Academy of Sciences, Beijing 100080, China}\\
$^b${\small Institute of Theoretical Physics, Shanxi University, Taiyuan\\
030006, China}\\
$^c${\small Department of Physics, University of Kaiserslautern, D-67653\\
Kaiserslautern, Germany}\\
$^d${\small Department of Physics, Guangzhou Normal College, Guangzhou\\
510400, China\bigskip}}
\maketitle

\begin{abstract}
{\bf Abstract} The quantum-classical transitions of the escape rates in a
uniaxial spin model relevant to the molecular magnet Mn$_{12}$Ac and a
biaxial anisotropic ferromagnetic particle are investigated by applying the
periodic instanton method. The effective free energies are expanded around
the top of the potential barrier in analogy to Landau theory of phase
transitions. We show that the first-order transitions occur below the
critical external magnetic field $h_x=\frac 14$ for the uniaxial spin model
and beyond the critical anisotropy constant ratio $\lambda =\frac 12$ for
the biaxial ferromagnetic grains, which are in good agreement with earlier
works.\bigskip

{\bf PACS number(s)}: 75.45.+j, 75.50.Tt

\bigskip
\end{abstract}

Macroscopic quantum tunneling(MQT) of magnetization was intensively studied
both theoretically and experimentally in the last decade\cite{Gunther}. In
recent years much attention has been attracted to the phase transition
problem in quantum tunneling of magnetization\cite{ChPRA,Chudnovsky1,Liang2}%
. The interest to this problem arises mainly from the successful experiments
on the molecular magnet Mn$_{12}$Ac\cite{Friedman,Hernadez,Thomas}. It has
been shown that\cite{ChPRA} the quantum-classical transition of the escape
rate is analogous to the phase transition and the general conditions for the
first- and second-order transitions are also analyzed. Quite recently an
effective free energy $F=a\phi ^2+b\phi ^4+c\phi ^6$ for the transitions of
a spin system was introduced\cite{Chudnovsky1} as in Landau theory of phase
transition. Here $a=0$ corresponds to the quantum-classical transition and $%
b=0$ to the boundary between first- and second-order transitions. In the
common sense the first-order transitions are difficult to find in real
systems and a uniaxial spin model is one of the very few examples which
would exhibit the first-order transition. Various theoretical methods were
used to deal with the spin tunneling problem\cite
{Enz,Sharf,Zaslavskii,Hemmen,Garanin,Chudnovsky2}. We present in this letter
the periodic instanton calculations for the quantum-classical transitions of
the escape rates for the uniaxial spin model and a biaxial anisotropy
ferromagnetic particle. The latter provides another candidate in which we
may observe the first-order transition at the proper anisotropy constant.

A rather simple and experimentally important system which may exhibit the
first-order transition is the uniaxial spin model in a magnetic field
parallel to $x$-axis $H_x$ described by the Hamiltonian 
\begin{equation}
{\cal H}=-DS_z^2-H_xS_x  \label{1}
\end{equation}
which is generic for problems of spin tunneling. This model is believed to
have relevance to the molecular magnet Mn$_{12}$Ac with $D$ the anisotropy
constant. It was found recently that there exists a critical value of the
external field below which the first-order escape rate transition occurs\cite
{Chudnovsky1}. Here the periodic instanton method is used to deal with this
problem which should be more accurate as we need not resort to the double
well potential approximation. Using the method of mapping the spin model
onto a particle problem\cite{Chudnovsky1,Sharf}, the equivalent particle
Hamiltonian is ${\cal H}=\frac{p^2}{2m}+U(x),$ where 
\begin{equation}
U(x)=S(S+1)D(h_x\cosh x-1)^2  \label{2}
\end{equation}
and $m=1/2D,$ $h_x=H_x/2\tilde{S}D,$ $\tilde{S}=S+1/2.$ The minimum of the
effective potential, $x_0=$ $\cosh ^{-1}(1/h_x),$ has been moved to zero and 
$S\gg 1$ is used throughout. Recently the single kink and kink lattice
solutions for a class of quasi-exactly solvable potential model in field
theory, the Double sinh-Gordon(DSHG) potential\cite{Khare}, have been found
and the statistical mechanics of DSHG kinks are studied. Our motivation is
based on that the potential (\ref{2}) is a special case of DSHG model and
corresponding instanton and periodic instanton configurations may be found
following the similar procedure.

\smallskip The vacuum instanton solution is nothing but the zero-energy
solution of the equation of motion in imaginary time $\tau $ 
\begin{equation}
\frac 12m\left( \frac{dx}{d\tau }\right) ^2-U(x)=0.  \label{3}
\end{equation}
The instanton/anti-instanton located at $\tau _0$ is given by 
\begin{equation}
x_c(\tau )=\pm 2\tanh ^{-1}\left[ \tanh \frac{x_0}2\tanh \left( \frac{\tau
-\tau _0}\xi \right) \right] ,\qquad \xi =\frac 1{\sqrt{1-h_x^2}\tilde{S}D}
\label{4}
\end{equation}
where $\tanh \frac{x_0}2=\sqrt{\frac{1-h_x}{1+h_x}}$ and the corresponding
Euclidean action is 
\begin{equation}
S_c=\int_{-\infty }^{+\infty }m\dot{x}_c^2d\tau =2\tilde{S}\left( \ln \frac{%
1+\sqrt{1-h_x^2}}{h_x}-\sqrt{1-h_x^2}\right) .  \label{5}
\end{equation}

To understand the finite temperature tunneling behavior we construct the
periodic instanton configuration satisfying the periodic boundary condition
similar to the kink lattice method in double sinh-Gordon(DSHG) theory. This
solution is obtained by integrating the equation of motion 
\begin{equation}
\frac 12m\left( \frac{dx}{d\tau }\right) ^2-U(x)=-E  \label{6}
\end{equation}
and is given by 
\begin{equation}
x_p(\tau )=\pm 2\tanh ^{-1}\left[ \tanh x_1%
\mathop{\rm sn}%
\left( \frac{\tau -\tau _0}{\xi _P},k\right) \right]  \label{7}
\end{equation}
with 
\begin{eqnarray}
k &=&\sqrt{\frac{1-\left( \sqrt{E^{\prime }}+h_x\right) ^2}{1-\left( \sqrt{%
E^{\prime }}-h_x\right) ^2}}=\frac{\tanh x_1}{\tanh x_2},  \label{8} \\
\xi _P &=&\frac 1{\tilde{S}D\sqrt{1-\left( h_x-\sqrt{E^{\prime }}\right) ^2}}%
=\frac k{2h_x\tilde{S}D\sinh x_1\cosh x_2}.  \nonumber
\end{eqnarray}
where $%
\mathop{\rm sn}%
\left( \tau ,k\right) $ is the Jacobi elliptic function with modulus $k$ and
the complementary modulus $k^{\prime }=\sqrt{1-k^2}$. Equivalently, 
\begin{equation}
\tanh ^2x_{1,2}=\frac{1-h_x\mp \sqrt{E^{\prime }}}{1+h_x\mp \sqrt{E^{\prime }%
}}=\frac{1-h_x}{1+h_x}\frac{\allowbreak (1+k^2)h_x\mp k^{\prime 2}(1+h_x)+%
\sqrt{4h_x^2k^2+k^{\prime 4}}}{(1+k^2)h_x\mp k^{\prime 2}(1-h_x)+\sqrt{%
4h_x^2k^2+k^{\prime 4}}}  \label{9}
\end{equation}
and the characteristic length of the periodic instanton 
\begin{equation}
\xi _P^2=\frac 1{\tilde{S}^2D^2}\left( 1-h_x^2\left( \frac{%
(1+k^2)h_x^2-k^{\prime 2}(1-h_x^2)+h_x\sqrt{4h_x^2k^2+k^{\prime 4}}}{\left(
1+k^2\right) h_x^2+h_x\sqrt{4h_x^2k^2+k^{\prime 4}}}\right) ^2\right) ^{-1}.
\label{10}
\end{equation}
This description corresponds to the movement of a pseudo-particle in the
inverted potential $-U(x)$ with energy $-E$ and $E^{\prime }=E/\tilde{S}^2D.$
The periodicity of the solution (\ref{7}) is $\tau _p(E)=4\beta ,$ $\beta =%
{\rm K}(k)\xi _P,$ where ${\rm K}(k)$ is the complete elliptic integral of
the first kind. The topological charge of periodic instanton 
\begin{equation}
Q_p=2x_p\left( {\rm K}(k)\right) =2x_1  \label{11}
\end{equation}
is smaller than the vacuum instanton case 
\begin{equation}
Q=\int_{-\infty }^{+\infty }\frac{\partial x_c(\tau )}{\partial \tau }%
dx=2x_0.  \label{12}
\end{equation}
Similarly, the periodic instanton size $\xi _P$ is also smaller than the
zero-energy one $\xi .$ The potential and the instanton configurations are
depicted in Fig. 1. The particle starts from the turning point $-x_1$ at
imaginary time $-\beta $ and reaches the other turning points $x_1$ at $%
\beta .$ After the same interval, at time $2\beta ,$ the particle returns to
its original position, i.e., it tunnels through the barrier twice in the
whole period.

The Euclidean action of the periodic instanton configuration is 
\begin{equation}
S_p=\int_{-\beta }^\beta \left[ m\stackrel{.}{x}_p^2+E\right] d\tau
=W+2E\beta  \label{13}
\end{equation}
where 
\begin{equation}
W=\frac 2{D\xi _P\alpha ^2}{\LARGE [}\left( \alpha ^4-k^2\right) {\rm \Pi }%
(\alpha ^2,k)+k^2{\rm K}(k)+\alpha ^2\left( {\rm K}(k)-{\rm E}(k)\right) 
{\LARGE ]}  \label{14}
\end{equation}
Here $\alpha ^2=\tanh ^2x_1<k^2$ and ${\rm E}(k),$ ${\rm \Pi }(\alpha ^2,k)$
are the complete elliptic integral of second and third kind, respectively.

The temperature dependent tunneling rate can be estimated by a Boltzmann
average over the tunneling probabilities from excited states with energy $E$
which is approximated by the semiclassical WKB exponents, $\Gamma _n\sim
e^{-2W(E)}.$ In the quasi-classical approximation the transition rate is
given by $\Gamma \sim e^{-F_{\min }/T},$ where $F_{\min }$ is the minimum of
the effective ''free energy''\cite{Chudnovsky1} 
\begin{equation}
F=E+TW^{\prime }(E),\qquad W^{\prime }(E)=2W(E)  \label{15}
\end{equation}
with respect to $E$. In the case of second-order transition the crossover
temperature is given by 
\begin{equation}
T_0^{(2)}=\frac{\widetilde{\omega }_0}{2\pi },\qquad \widetilde{\omega }_0=%
\frac{\tilde{S}D}\pi \sqrt{h_x\left( 1-h_x\right) }  \label{16}
\end{equation}
where $\widetilde{\omega }_0=\sqrt{\left| U^{\prime \prime }(0)\right| /m}$
is the instanton frequency. It is convenient to introduce dimensionless
temperature and energy variables 
\begin{equation}
\theta =\frac T{T_0^{(2)}},\qquad P=\frac{\Delta U-E}{\Delta U},  \label{17}
\end{equation}
where $\Delta U=U_{\max }-U_{\min }$ is the barrier height. To investigate
the phase transition behavior, we need to expand the free energy around the
top of the potential barrier. Near the potential maximum($k\sim 0$) the
expansion of elliptic integrals up to order of $k^6$ is seen to be\cite{Byrd}
\begin{eqnarray}
{\rm K}(k) &=&\frac \pi 2\left[ 1+\frac{k^2}4+\frac 9{64}k^4+\frac{25}{256}%
k^6+\cdots \right]  \label{18} \\
{\rm E}(k) &=&\frac \pi 2\left[ 1-\frac{k^2}4-\frac 3{64}k^4-\frac 5{256}%
k^6-\cdots \right]  \nonumber \\
{\rm \Pi }(\tanh ^2x_1,k) &=&\frac \pi 2+\frac \pi 8(3-2h_x)k^2+\frac \pi {%
128}\left( 32h_x^3-8h_x^2-60h_x+45\right) k^4  \nonumber \\
&&+\frac \pi {512}\left( -\allowbreak
256h_x^5+64h_x^4+480h_x^3-88h_x^2-350h_x+175\right) k^6.  \nonumber
\end{eqnarray}
The other parameters in the free energy, such as $\alpha ^2$ and $\xi _p,$
are also calculated in the same way and we obtain the result 
\begin{eqnarray}
F/\Delta U &=&1+4h_x\left( \theta -1\right) k^2+4h_x\left(
h_x^2+2h_x-1-\theta \left( h_x^2+\frac 32h_x-\frac 78\right) \right) k^4 
\nonumber \\
&&+4h_x\left( \theta \left( 2h_x^4+3h_x^3-h_x^2-\frac{11}4h_x+\frac{51}{64}%
\right) -\left( 2h_x^4+4h_x^3-4h_x+1\right) \right) k^6.  \label{19}
\end{eqnarray}
There exists an exact relation between $k^2$ and $P$%
\begin{equation}
k^2=\frac{-2h_x-P+h_xP+2h_x\sqrt{1-P}}{-2h_x-P+h_xP-2h_x\sqrt{1-P}}.
\label{20}
\end{equation}
Expressing $k^2$ in power series of $P,$ we have 
\begin{equation}
F/\Delta U=1+(\theta -1)P+\frac \theta 8(1-\frac 1{4h_x})P^2+\frac{3\theta }{%
64}(1-\frac 1{3h_x}+\frac 1{16h_x^2})P^3+O(P^4)  \label{21}
\end{equation}
which coincides with Ref. \cite{Chudnovsky1} exactly. There indeed exists a
phase boundary between the first- and second-order transitions, i.e., $h_x=%
\frac 14,$ at which the factor in front of $P^2$ changes the sign. We
conclude that this critical boundary is inherent in the DSHG model and plays
dominant role in the tunneling process of the uniaxial spin system with an
external magnetic field, eq. (1).

Turning now to the computation of level splittings of excited states. We
have a more general formula for the double- well-like potentials in WKB
approximation\cite{Haeffner,Shepard,Chudnovsky1} 
\begin{equation}
\Delta E=\frac{\omega (E)}\pi \exp [-W]  \label{22}
\end{equation}
where $\omega (E)=\frac{2\pi }{t(E)}$ is the energy-dependent frequency and $%
t(E)$ is the period of the real-time oscillation in the potential well. This
level splitting formula is valid for entire energy region $0<E<\Delta U.$
The calculation of the period $t(E),$ equivalently the normalization
constant of WKB wave functions\cite{Liang3}, results in 
\begin{equation}
t(E)=\sqrt{2m}\int_{x_1}^{x_2}\frac{dx}{\sqrt{E-U(x)}}=\frac{2{\rm K}%
(k^{\prime })}{\tilde{S}D\sqrt{1-\left( \sqrt{E^{\prime }}-h_x\right) ^2}}.
\label{23}
\end{equation}
For energies near the bottom of the potential well the energy-dependent
frequency tends to the classical frequency of small oscillations at the
bottom of the well $\omega _0=2\tilde{S}D\sqrt{1-h_x^2}$ while near the
barrier top this frequency reduces to the instanton frequency $\widetilde{%
\omega }_0$. Here we discuss the low energy limit of the level splitting.
Under the condition $E\ll \Delta U,$ the action $W$ may be expanded in power
series of $k^{\prime }$ according to another group of formulae 
\begin{eqnarray}
{\rm K}(k) &=&\ln \frac 4{k^{\prime }}+\frac 14\left( \ln \frac 4{k^{\prime }%
}-1\right) k^{\prime 2}+\frac 9{64}\left( \ln \frac 4{k^{\prime }}-\frac 76%
\right) k^{\prime 4}  \label{24} \\
{\rm E}(k) &=&1+\frac 12\left( \ln \frac 4{k^{\prime }}-\frac 12\right)
k^{\prime 2}+\frac 3{16}\left( \ln \frac 4{k^{\prime }}-\frac{13}{12}\right)
k^{\prime 4}  \nonumber \\
{\rm \Pi }(\tanh ^2x_1,k) &=&-\frac{\sqrt{1-h_x^2}}{2h_x}x_0+\frac{1+h_x}{%
2h_x}\ln \frac 4{k^{\prime }}-\frac{1+h_x+\sqrt{1-h_x^2}x_0}{8h_x}k^{\prime
2}+\frac{h_x+1}{8h_x}k^{\prime 2}\ln \frac 4{k^{\prime }}  \nonumber \\
&&-\left( \frac{4h_x^2-1}{64h_x^3}\sqrt{1-h_x^2}x_0+\frac 3{256}\frac{%
7h_x^3+8h_x^2-1}{h_x^3}\right) k^{\prime 4}+\frac{9h_x^3+11h_x^2-2}{128h_x^3}%
k^{\prime 4}\ln \frac 4{k^{\prime }}  \nonumber
\end{eqnarray}
and $k^{\prime 4}\approx 16E^{\prime }h_x^2/(1-h_x^2)^2.$ Inserting them
back into (\ref{14}) we obtain 
\begin{equation}
W=2\tilde{S}\sqrt{1-h_x^2}\left( \frac{2x_0}{\sqrt{1-h_x^2}}-1-\frac{1-h_x^2%
}{64h_x^2}k^{\prime 4}-\frac{1-h_x^2}{16h_x^2}k^{\prime 4}\ln \frac 4{%
k^{\prime }}\right)  \label{25}
\end{equation}
which reduces to the ground state action (\ref{5}) at $k^{\prime }=0.$ This
expression may be rewritten in a more compact form 
\begin{equation}
W=S_c(0)-\frac E{\omega _0}\ln \frac{eq\omega _0}E  \label{26}
\end{equation}
with 
\begin{equation}
q=\frac{8\tilde{S}\left( 1-h_x^2\right) ^{3/2}}{h_x^2}.  \label{27}
\end{equation}
Approximating the energy levels in the well by a harmonic oscillator, i.e., $%
E=(n+\frac 12)\omega _0,$ and taking into account corrections from the
functional-integral technique\cite{Haeffner,Shepard,Chudnovsky1}, simplify (%
\ref{26}) into 
\begin{equation}
W=S_c(0)-\ln \left( \frac{8\tilde{S}e\left( 1-h_x^2\right) ^{\frac 32}}{%
\left( n+\frac 12\right) h_x^2}\right) ^{(n+\frac 12)},  \label{28}
\end{equation}
so the low-lying energy level splitting takes the form\cite
{Haeffner,Zaslavskii} 
\begin{equation}
\Delta E_n=\frac{\Delta E_0}{n!}q^n  \label{29}
\end{equation}
where the ground state splitting 
\begin{equation}
\Delta E_0=\frac{8\tilde{S}^{\frac 32}D}{\pi ^{\frac 12}}\left[ \frac{\exp 
\sqrt{1-h_x^2}}{1+\sqrt{1-h_x^2}}\right] ^{2\tilde{S}}\left( 1-h_x^2\right)
^{5/4}h_x^{2\tilde{S}-1}\sim h_x^{2S}  \label{30}
\end{equation}
is proportional to power $2S$ of the perturbative transverse field $h_x$\cite
{Zaslavskii}.

Now consider the phase transition in a biaxial anisotropic ferromagnetic
grain without applied magnetic field 
\begin{equation}
{\cal H}=K_1S_z^2+K_2S_y^2  \label{31}
\end{equation}
which has been investigated intensively\cite{Enz,Chudnovsky2,Liang1}. This
model possesses a XOY easy plane and an easy axis along the $x$ direction
with the anisotropy constants $K_1>K_2>0$. We show here that it provides
another example which exists first-order transition from classical to
quantum regimes. With the help of the coherent--state path integral the
effective Hamiltonian in terms of continuum field variable can be written in
the form ${\cal H}=\frac{p^2}{2m(\phi )}+U(\phi )$ and 
\begin{equation}
m(\phi )=\frac 1{2K_1(1-\lambda \sin ^2\phi )},\quad U(\phi )=K_2S(S+1)\sin
^2\phi  \label{32}
\end{equation}
where $\lambda =\frac{K_2}{K_1}<1$ and mass $m(\phi )$ is field dependent.
The periodic instanton solution is given by\cite{Liang1} 
\begin{equation}
\phi _p=\arcsin \sqrt{\frac{1-k^2%
\mathop{\rm sn}%
^2\left( \omega \tau ,k\right) }{1-\lambda k^2%
\mathop{\rm sn}%
^2\left( \omega \tau ,k\right) }}  \label{33}
\end{equation}
with 
\begin{equation}
k=\sqrt{\frac{n_1^2-1}{n_1^2-\lambda }},\qquad n_1^2=\frac{K_2S(S+1)}{E_{cl}}
\label{34}
\end{equation}
where $\omega =\omega _0\sqrt{1-\lambda /n_1^2}$ denotes the small
oscillation frequency at the position of periodic instanton and $\omega
_0^2=4S(S+1)K_1K_2$. The non-monotonically decreasing behavior of the
periodicity of the solution, $\tau (E)=4{\rm K}(k)/\omega ,$ has been
pointed in Ref. \cite{Liang2} where the authors proved that beyond a
critical value of coupling the spin system acquired a first order transition
as a result of the field dependence of its effective mass. We now turn to
evaluate the effective free energy dependence on dimensionless energy scale $%
P$ for various anisotropy constants.

The Euclidean action evaluated for the periodic instanton trajectory is
given by eq. (\ref{13}) with 
\begin{equation}
W=\frac \omega {\lambda K_1}\left[ {\rm K}(k)-(1-\lambda k^2){\rm \Pi }%
(\lambda k^2,k)\right] .  \label{35}
\end{equation}
The second order transition temperature for this model is defined by eq.(\ref
{16}) with the instanton frequency $\widetilde{\omega }_0=\sqrt{\left|
U^{\prime \prime }(\pi /2)\right| /m(\pi /2)}=\sqrt{1-\lambda }\omega _0.$
Near the top of the barrier the third kind of elliptic integral should be
expanded as 
\begin{equation}
\Pi (\lambda k^2,k)=\frac \pi 2+(2\lambda +\allowbreak 1)\frac \pi 8%
k^2+\left( 8\lambda ^2+4\lambda +3\right) \frac{3\pi }{128}k^4+\left(
16\lambda ^3+8\lambda ^2+6\lambda +5\right) \frac 5{512}\pi k^6.  \label{36}
\end{equation}
After a straightforward calculation we obtain the expansion result for the
free energy (\ref{15}) in terms of $k$ up to the sixth order 
\begin{equation}
F/\Delta U=1-P+\theta \left( 1-\lambda \right) \left( k^2+\left( 6\lambda
+1\right) \frac{k^4}8+\left( 40\lambda ^2+8\lambda +3\right) \frac{k^6}{64}%
\right) .  \label{37}
\end{equation}
The effective free energy analogous to Landau theory of phase transitions
near the top of the barrier ($P\ll 1$) reads 
\begin{equation}
F/\Delta U=1+\left( \theta -1\right) P+\frac \theta {8\left( 1-\lambda
\right) }\left( 1-2\lambda \right) P^2+\frac \theta {64\left( 1-\lambda
\right) ^2}\left( 8\lambda ^2-8\lambda +3\right) P^3+O(P^4)  \label{38}
\end{equation}
with the exact relation between $k^2$ and $P$%
\begin{equation}
k^2=\frac P{1-\lambda (1-P)}.  \label{39}
\end{equation}
The factor before $P$ changes sign at the phase transition temperature $%
T=T_0^{(2)}.$ The boundary between the first- and the second-order
transition is obviously seen to be $\lambda _c=\frac 12.$ The computed
dependence of $F$ on $P$ for the entire range of energy is plotted in Fig. 2
for $S^2=1000,K_1=1$. At $\lambda =0.3,$ there is only one minimum of $F$ at
the top of the barrier for all $T>T_0^{(2)}.$ Below $T_0^{(2)}$ it
continuously shifts to the bottom as the temperature is lowered. This
corresponds to the second-order transition from thermal activation to
thermally assisted tunneling. At $\lambda =0.8,$ however, there can be one
or two minima of $F,$ depending on the temperature. The actual phase
transition(in this case the first-order) occur at the temperature when the
two minima have the same free energy, which for $\lambda =0.8$ takes place
at $T_0^{(1)}=1.122T_0^{(2)}.$

The relation between tunneling at excited states and finite temperature can
be understood as follows: Below the crossover temperature $T_0$ the particle
tunnels through the barrier at the most {\it favorable} energy level $E(T)$
which goes down from the top of the barrier to the bottom of the potential
with lowering temperature. Such a regime is called thermally assisted
tunneling(TAT). The second order transition from the classical thermal
activation to TAT is smooth, with no discontinuity of $d\Gamma /dt$ at $T_0,$
and the transition temperature is given by $T_0^{(2)}.$ In the situation
under which the first order transition may occur tunneling just below the
top of the barrier is {\it unfavorable}, the TAT is suppressed, and the
thermal activation competes with the ground state tunneling directly,
leading to the first order transition.

The flatness of the barrier top is not the necessary condition under which
the first order transition may occur. For the constant mass model the more
favorable potential for the first order transition is that the top of the
barrier should be wider so that the particle doesn't have more advantage to
tunnel through the barrier from the excited states than from the ground
state. In our second model the pseudo-particles near the top of the barrier
are ''heavier'' than those in the bottom of the well, i.e. $m(\pi
/2)=m(0)/(1-\lambda )>m(0)$. The advantage for the particles to tunnel
through the barrier at higher excited states is again not very obvious,
which leads to the first order transition from the thermal activation
directly to ground state tunneling.

A simple estimation for the crossover temperature $T_0^{(0)}$ is given by 
\begin{equation}
T_0^{(0)}=\frac{\Delta U}{2S_c(E=0)}=\frac{K_2(S+\frac 12)}{2\ln \frac{1+%
\sqrt{\lambda }}{1-\sqrt{\lambda }}}  \label{40}
\end{equation}
where the superscript of $T_0^{(0)}$ indicates that the ground state
tunneling is considered. In Fig. 3 we plot the dependence of the actual
crossover temperature on $\lambda $. For $\lambda <\frac 12,$ the actual
crossover temperature $T_0^{(1)}$ is read out from Fig. 2(b). The escape
rate can be conveniently represented in terms of the effective temperature
defined by 
\begin{equation}
\Gamma \sim \exp [-\frac{\Delta U}{T_{eff}(T)}]=\exp [-\frac{F_{\min }}T]
\label{41}
\end{equation}
The dependence $T_{eff}(T)=\Delta UT/F_{\min }$ is represented in Fig. 4 for
different anisotropy constant ratio $\lambda .$ The most significant
difference between the crossover temperature $T_0^{(0)}$ and the actual
crossover temperature $T_0$ arises in the limit $\lambda \rightarrow 0,$
that is, $T_0^{(0)}/T_0^{(2)}=\pi /4\approx \allowbreak 0.\,785$.

The first-order escape-rate transition considered above is the transition
from thermal activation to thermally assisted tunneling near the bottom of
the potential and not directly to the ground-state tunneling\cite
{Chudnovsky1} due to the logarithmic divergence of the instanton period $%
\tau (E)$ for the energies near $U_{\min }$. In some field-theoretical
models, as, e.g., the reduced nonlinear $O(3)-$ $\sigma $ model, $\tau $
approaches $0$ near the bottom of the potential. Accordingly, the second
derivative of $W(E)$ and $F(E)$ is negative everywhere, as for the
rectangular potential for particles. In such a situation, as it is clear
from Fig. 2(b), the minimun of $F(E)$ can only be at the barrier top or
potential bottom. That is, thermal activation competes directly with the
ground-state tunneling. It was shown that\cite{Zimmer} adding a small Skyrme
term to the reduced nonlinear $O(3)-$ $\sigma $ model causes $\tau $ to
diverge near the bottom of the potential, with the accordingly small
amplitude. This is, in a sense, similar to the situation realized in this
spin model for $\lambda \rightarrow 1$.

In conclusion, we present in this letter the periodic instanton calculation
for the first- and second-order transitions between quantum and classical
regimes for two spin models. The level splitting formula for the excited
states is derived and checked for the uniaxial spin system$.$ Our results
for uniaxial spin model confirm Ref. \cite{Chudnovsky1} and for the biaxial
anisotropic spin system we find that the transition from the classical
regimes with lowering temperature is of the first order for $\lambda $ above
the phase boundary line $\lambda _c=\frac 12,$ and of the second-order below
this critical value.

Helpful discussions with Dr. X. B. Wang, Prof. Lee Chang and Prof. B. Z. Li
are appreciated. This Work is partly supported by the National Natural
Science Foundation of China under Grant Nos. 19677101 and 19775033.

{\bf Figure Captions:}

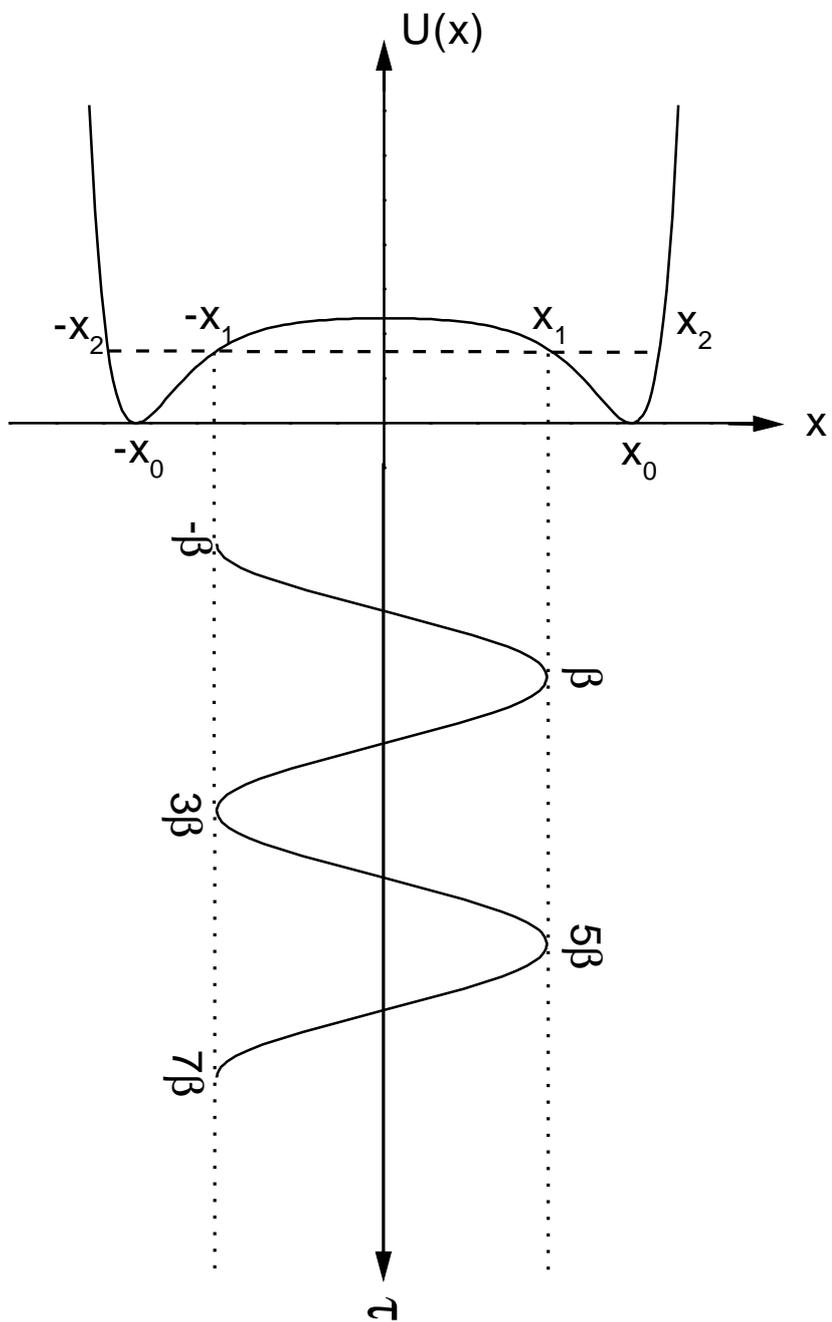
\smallskip Fig. 1. The DSHG potential and the periodic instanton
configuration for $S=10,D=0.6K,h_x=0.01$

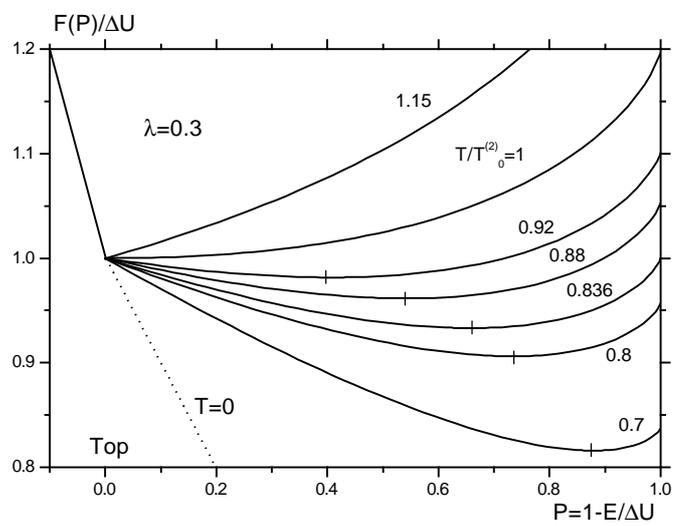
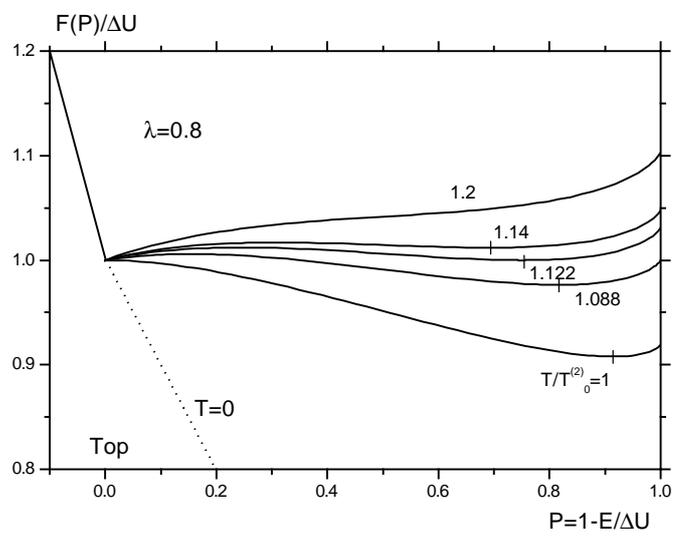
Fig. 2. Effective free energy for the escape rate: (a) $\lambda =0.3$,
second-order transition; (b) $\lambda =0.8$, first order-transition.

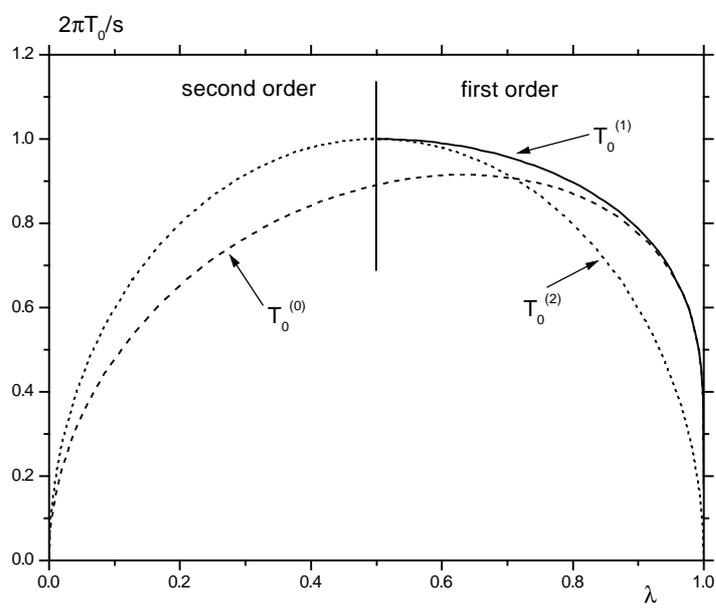
Fig. 3. Dependence of the actual phase transition temperature $T_0$ on the
anisotropy constant ratio.

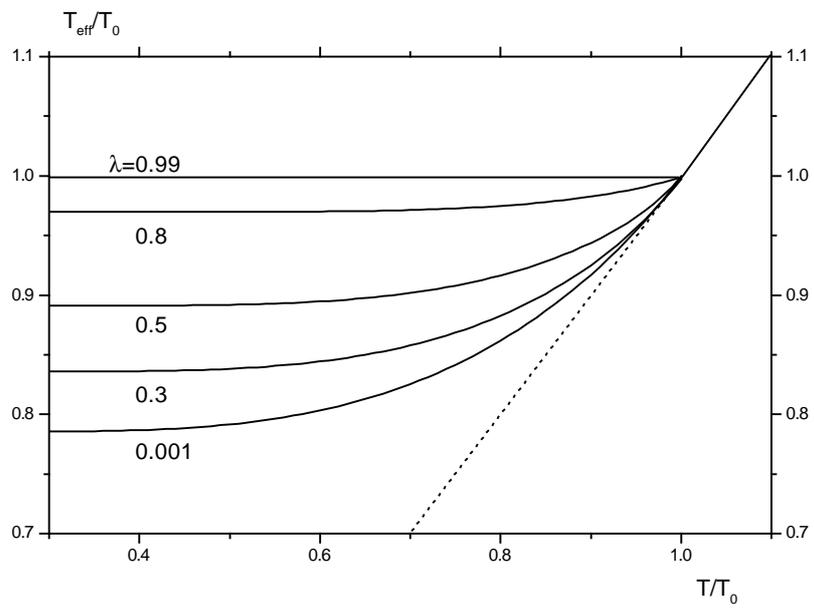
Fig. 4. Dependence of the effective temperature $T_{eff}$ on $T$ for
different values of $\lambda $

\end{document}